\documentclass[12pt]{article}
\usepackage[utf8]{inputenc}

\usepackage{graphicx}
\usepackage{natbib}
\usepackage{amsmath}
\usepackage{amssymb}
\usepackage[margin=1in]{geometry}
\usepackage[multiple]{footmisc}
\usepackage{url}
\usepackage{hyperref}

\title{What we cannot learn from analogue experiments\thanks{We thank audiences in Bad Honnef, Seven Pines, Geneva, Salzburg, Dubrovnik, and Bern, as well as Radin Dardashti, Karim Th\'ebault, and Marcel Weber for discussions. We acknowledge financial support (for K.C. and N.L.) from the Swiss National Science Foundation (Project $105212\_165702$) and (for C.W.) from the John Templeton Foundation (Grant 56314, made under a collaborative agreement between the University of Illinois at Chicago and the University of Geneva). The content of this work are solely the responsibility of the authors and do not represent the official views of the Swiss National Science Foundation or the John Templeton Foundation.}}
\author{Karen Crowther, Niels Linnemann, Christian W\"uthrich\thanks{Department of Philosophy, University of Geneva}}

\date{29 March 2019}

\begin{document}

\maketitle

\begin{abstract}\noindent
Analogue experiments have attracted interest for their potential to shed light on inaccessible domains. For instance, `dumb holes' in fluids and Bose-Einstein condensates, as analogues of black holes, have been promoted as means of confirming the existence of Hawking radiation in real black holes. We compare analogue experiments with other cases of experiment and simulation in physics. We argue---\textit{contra} recent claims in the philosophical literature---that analogue experiments are not capable of confirming the existence of particular phenomena in inaccessible target systems. As they must assume the physical adequacy of the modelling framework used to describe the inaccessible target system, arguments to the conclusion that analogue experiments can yield confirmation for phenomena in those target systems, such as Hawking radiation in black holes, beg the question.

\vspace{3mm}
\noindent \textbf{Keywords:} Analogy; Black Hole Thermodynamics; Hawking Radiation; Dumb Holes; Quantum Gravity; Analogue Reasoning.
\end{abstract}

\tableofcontents

\section{Introduction: analogue gravity}

Quantum gravity---the field of physics centred around the (currently unknown) theory that is required in order to describe the domains where both quantum and general relativistic effects are thought to be non-trivial---is notorious for its lack of experimental or observational data. Thus, the development of theories in this domain turns on theoretical considerations, contended over in an empirical vacuum. In this context, black holes have been hailed as the Rosetta Stone which supposedly furnishes the combination of relativistic and quantum effects that will lead the way to quantum gravity. This hope is based on results in black hole thermodynamics, which rests on a suggestive analogy between the laws of ordinary thermodynamics and the so-called `black hole thermodynamic laws' \citep[206]{KieferBook}. The analogy relates quantities such as surface gravity and horizon area defined in terms of the basic properties of black holes (their mass, charge, and angular momentum) to thermodynamic properties such as temperature and entropy. When \citet{bekenstein1972,bekenstein1973} first proposed the analogy, it was not taken seriously---for good reason: a classical black hole as it is described in general relativity is a perfect absorber and must thus be of zero temperature, despite its non-zero surface curvature. Consequently, the analogy was initially taken as a mere formal curiosity with no physical meaning.\footnote{\citeauthor{Wuthrichbekenstein} (Forthcoming) offers an analysis of Bekenstein's argument and an assessment of the strength of the analogy he argues for, arriving at a largely negative conclusion concerning Bekenstein's original case. \citeauthor{DoughertyCallender} (Forthcoming) and \citet{WallaceThermodynamics} undertake a more general evaluation of the strength of the analogy, and reach opposing conclusions.} 

The situation drastically changed when \citet{hawking1975} showed how semi-classical considerations, i.e.\ calculations based partly on relativistic and partly on quantum physics, can lead to a form of thermal radiation being apparently emitted from black holes. The existence of this `Hawking radiation' made an assignment of a genuine physical temperature $T_H$ to black holes meaningful. Strikingly, it turns out that $T_H$ is  proportional to the black hole's surface gravity, $\kappa$, and so fits the role of temperature in the black hole thermodynamic laws: as $dS \propto dA$ in the analogy---i.e., (infinitesimal) entropy, $dS$, is proportional to the (infinitesimal) black hole horizon area, $dA$---, $T_H \propto \kappa$ allows for the right relationship between temperature and entropy, that is, $T dS = \frac{\kappa}{8 \pi G} dA$ (where $G$ is Newton's gravitational constant). Thus, now following Bekenstein, most physicists seem to have accepted that the association between black holes and thermodynamic systems is not just one of a physically meaningful analogy; instead, they conclude that black holes are thermodynamic objects, and thus that a black hole's surface gravity is a temperature and a black hole's horizon surface area is an entropy of the same kind as that exhibited by any terrestrial thermodynamic system (see for instance \cite[\S1 et passim]{WallaceThermodynamics} for an account on the reception of black hole thermodynamics).\footnote{Note, however, that physicists are premature in accepting the hypothetical temperature of Hawking radiation as being an actual temperature---and thereby in promoting the analogy between thermodynamics and black holes to an identity.} In fact, it is the identification of a black hole as truly thermodynamic that permits the generalisation of the statement that entropy never decreases for a closed system (the second law of thermodynamics) to the statement that the sum of the ordinary entropy and the black hole entropy never decreases for a closed system (now known as the \textit{generalised second law}). 

Although physicists are confident of the association between horizon area and entropy, and thus between black holes and thermodynamics, it stands and falls with the actual existence of Hawking radiation in black holes---currently, like the rest of black hole thermodynamics, only a theoretical hypothesis. Thus, an empirical confirmation of Hawking radiation would strongly support the belief that the analogy actually holds in the physical world, and so constitute an important stepping stone to an empirical access to quantum gravity. Although astrophysical black holes (i.e., black holes which are not `small') are being routinely (indirectly) detected today, their potential Hawking radiation is far too weak to be detectable against the fluctuations of the cosmic microwave background. Furthermore, black holes are evidently not the kind of systems one can easily manipulate in ordinary laboratory practice---they are systematically inaccessible to our standard experimental treatment.\footnote{We will use the term `inaccessible' in a non-technical sense throughout, hoping that, whatever its precise characterisation, it should be clear that astrophysical black holes are experimentally inaccessible.} In an attempt to overcome these difficulties, \citet{unruh1981} proposed the study of `analogue' systems---such as `dumb holes', where sound waves in fluids encounter a horizon---and argued that from these studies, we can learn something about black holes. In what turned out to be the birth of what is now the thriving field of `analogue gravity', Unruh relied on the basic idea of \textit{analogical reasoning}: if two systems are known to be similar in some relevant respects, then we are justified in thinking that they are also similar in some unknown respects of interest. Concretely, the argument trades on the idea that if black holes and analogue systems in the lab are formally described by models that are relevantly the same, then either both or neither should exhibit Hawking radiation. Recently, \citet{Steinhauer} has reported the detection of (spontaneous) Hawking radiation in experiments where Bose-Einstein condensates served as analogue systems.\footnote{This is not the only work on this: for example, \cite{rousseaux2008observation}, \cite{Rousseaux2010horizon}, \cite{Weinfurtner} and \cite{French} (among others) had indeed already found the analogue Hawking effect in fluid systems (using water tanks). These analogue systems can however only mimic \textit{stimulated} Hawking radiation (sometimes referred to as the `classical Hawking effect') as opposed to \textit{spontaneous} Hawking radiation (as in Steinhauer's experiment). As gravitational Hawking radiation can only be spontaneous, Steinhauer's finding allows for the stronger analogy.}

Thus, the crucial question arises as to what extent analogue experiments such as those proposed by Unruh and conducted by Steinhauer can shed light on black holes. To date, philosophical examination of analogue experiments has mainly focused on the question of whether or not they are capable of providing \textit{confirmation}---that is, whether analogue experiments can give us reason to believe in (or, quantitatively, increase our credence in) particular hypotheses about the systems they are supposed to represent. This paper offers an analysis of precisely this issue, reaching a conclusion opposed to the affirmative one argued for in the philosophical literature by Dardashti, Th\'ebault, and Winsberg (DTW, \citet{DardashtiQualitative}), who articulate specific conditions under which analogue experiments could be confirmatory, reaffirmed by \citeauthor{Thebault2016} (Forthcoming), and again by  Dardashti, Hartmann, Th\'ebault, and Winsberg (DHTW, \citet{DardashtiHawkingQuantitative}) using a Bayesian analysis to quantify the potential degree of confirmation that analogue experiments could provide. A secondary topic of discussion in the literature has been whether or not analogue experiments differ from more familiar (and less concretely manipulable) cases of analogue reasoning. DTW (2017) and \citeauthor{Thebault2016} (Forthcoming) each argue that analogue experiment is distinct from analogy, and that unlike mere analogies---which are standardly taken to lack any confirmatory power, analogue experiments can (under specific circumstances) provide confirmation.

Our main thesis is that analogue experiments do not yield either confirmation or disconfirmation of hypotheses concerning black holes, such as that they radiate as predicted by Hawking. In support of this thesis, we argue that DTW's (2017) arguments for showing the potential of analogue experiments to provide confirmation of the existence of some phenomenon in an inaccessible target system beg the question, and thus do not establish their intended conclusion.\footnote{\label{fn:fallacy}Since the confirmation these authors are seeking is of course inductive rather than deductive, we do not accuse them of committing the fallacy of deductive reasoning known under the same name. However, we do accuse them of an inductive analogue of that fallacy and thus use the same expression.} In particular, these arguments applied to the case of dumb holes do not show that such experiments can confirm the existence of Hawking radiation in black holes. Concretely, the alleged confirmation of Hawking radiation crucially rests on the assumptions that quantum field theory (QFT) in curved spacetime is applicable to the case at hand, and that, specifically, the high-energy physics is independent of the trans-Planckian physics---assumptions which should be \textit{confirmed}, rather than be \textit{presupposed}, by a discovery of Hawking radiation. We thus claim that analogue models of Hawking radiation are no more or less confirmatory than the analogy between thermodynamics and black hole thermodynamics, for example.

In this paper, we are interested only in analogue experiments whose target systems (i.e., the systems they are supposed to represent) are inaccessible. This accords with the discussion of analogue experiments so far in the philosophical literature, where they are primarily of interest for their potential to penetrate the frontier of physics. As we argue in \S\ref{sec:analexp}, we take this inaccessibility to be the defining feature of analogue experiments. And, our major claim in this paper (as a whole) is that analogue experiments cannot tell us whether or not a particular phenomenon in the inaccessible target system exists. 

The plan of the paper is as follows. After introducing some basics of analogical reasoning, \S\ref{sec:analexp} asks what is special about analogue experiments, by comparing them with conventional experiments and simulation (\S\ref{ssec:CE}). Our primary examples of analogue experiments involve, of course, dumb holes. These are presented in \S\ref{sec:HR}, including a description of the derivation of Hawking radiation, a recap of the analogous derivation of the analogue hydrodynamic phenomenon, and a depiction of the recent experimental results on this matter, as for instance obtained by \cite{Steinhauer} in Bose-Einstein condensates. In \S\ref{sec:why}, we consider the conditions under which analogue experiments may be confirmatory, according to DTW (2017) and \citeauthor{Thebault2016} (Forthcoming). In \S\ref{sec:critic}, we show how DTW's argument for how analogue experiments may confirm the existence of a phenomenon in a target system already builds in the assumption that the target system displays the phenomenon in question. Conclusions follow in \S\ref{sec:conc}.

\section{What is an analogue experiment?}\label{sec:analexp}

We take the prime examples of analogue experiments in physics to be \textit{dumb holes}: laboratory experiments in fluid set-ups that are analogues of black holes. A simple example is the demonstration of an analogue black hole horizon using water draining from a bathtub and a speaker below the drain hole, transmitting sound waves up through the draining water. If the speed of the water as it drains out exceeds the speed of sound waves travelling in the opposite direction in the water, then no sound can escape from this point: one obtains an `acoustic horizon'. These systems are thus known as `dumb holes'---as in \textit{mute}, unable to speak. The target system is a black hole, which possesses a horizon delimiting a region of spacetime from which no light can escape, due to extreme curvature of spacetime. The water in the source system, $S$, is thus supposed to be analogous to spacetime in the target system, $T$. The salient properties of the source system are the speed of sound and the speed of the water, which are analogous to the speed of light, and the curvature of spacetime (i.e., gravity), respectively, in the target system. Other hypothesised black hole phenomena can be explored using similar analogue systems, including superradiance in rotating black holes \citep{superradiance}, and Hawking radiation \citep{Steinhauer}, the latter which we consider in more detail below (\S\ref{sec:HR}).

An analogue experiment is so-called because it relies on analogical reasoning.\footnote{We take `analogue' to be synonymous with `analogical'.} Following \cite{sep-reasoning-analogy}, an argument by analogy has the following form:
\begin{enumerate}
\item[P1.] $S$ is similar to $T$ in certain (known) respects.
\item[P2.] $S$ has some further feature $Q$.
\item[C.] Therefore, $T$ also has the feature $Q$, or some feature $Q^*$ similar to $Q$.
\end{enumerate}

\noindent Here, $S$ is the source system, and $T$ is the target system, i.e., the system of interest. The argument form is inductive rather than deductive, meaning that the conclusion is not guaranteed to follow from the premises. The consensus in the philosophical literature (see \cite{sep-reasoning-analogy} for a review) is that analogue reasoning---though certainly useful and ubiquitous in science---only establishes the \textit{plausibility} of a hypothesis. That is, analogies may confer some heuristic support for a hypothesis, providing reason for scientists to investigate it further.\footnote{See Bartha (2016, \S2.3), Bartha (2010, \S8.5), and, for a rather positive account on the linkage between confirmation and analogical argument, \cite{Hesse1964}.} According to this consensus, analogies are primarily of heuristic value: motivating additional research (e.g., experiments), rather than confirming hypotheses about $T$.

An analogy between $S$ and $T$ involves a mapping between parts of $S$ and those of $T$. Not all of the parts of $S$ and $T$ need to be placed in correspondence, however; usually, an analogy only relates some of the entities of the two systems, identifying the most significant similarities between $S$ and $T$. DTW (2017) point to a particular case of analogy, identified by \cite{Hempel1965}, where there is a \textit{syntactic isomorphism} between the laws describing $S$ and those describing $T$. \citet[][208f]{BarthaBook} says of this that the ``essential idea is that the two sets of physical laws have a common mathematical form and may be obtained by assigning different physical interpretations to the symbols that appear in that common form''. DTW (2017) adopt this idea of analogy as syntactic isomorphism, but generalise it so that it may apply more broadly---for instance, to cases where it is not the laws of $S$ and $T$ that are isomorphic, but the particular \textit{modelling frameworks} used to describe each of the systems, where these modelling frameworks have narrower scope than laws or theories do.

DTW then define an \textit{analogue simulation} in the following way (though, as we explain below, we think the term `analogue \textit{experiment}' would be more appropriate):

\paragraph*{}
\textbf{DTW's characterisation of an analogue simulation}\label{def:ae}

\begin{quote}
``[A] system $S$ provides an analogue simulation of system $T$ when the following set of conditions obtain [footnote suppressed]:
\begin{enumerate}
\item[Step 1:] For certain purposes and to a certain degree of desired accuracy, modelling framework $M_S$ is adequate for modelling system $S$ within a certain domain of conditions $D_S$.
\item[Step 2:] For certain purposes and to a certain degree of desired accuracy, modelling framework $M_T$ is adequate for modelling system $T$ within a certain domain of conditions $D_T$.
\item[Step 3:] There exists exploitable mathematical similarities between the structure of $M_S$ and $M_T$ sufficient to define a syntactic isomorphism robust within the domains $D_S$ and $D_T$.
\item[Step 4:] We are interested in knowing something about the behaviour of a system $T$ within the domain of conditions $D_T$, and to a degree of accuracy and for a purpose consistent with those specified in Step 2. For whatever reasons, however, we are unable to directly observe the behaviour of a system $T$ in those conditions to the degree of accuracy we require.
\item[Step 5:] We are, on the other hand, able to study a system $S$ after having put it under such conditions as will enable us to conclude a statement of the form:
\item[Claim\textsubscript{S}:] Under conditions $D_S$ and to degree of accuracy that will be needed below, we can for the purpose of employing the reasoning below assert that a system $S$ will exhibit phenomena $P_S$.
\end{enumerate}
The formal similarities mentioned in Step 3 then allow us to reason from Claim\textsubscript{S} to Claim\textsubscript{T}, which is of the form:
\begin{enumerate}
\item[Claim\textsubscript{T}:] Under conditions $D_T$, a system $T$ will exhibit phenomena $P_T$.''
\end{enumerate}
(DTW 2017, 67).
\end{quote}

This characterisation of an analogue experiment is also DTW's argument for how analogue experiments can be confirmatory---we return to the issue of confirmation in \S4--5. For now, however, we focus on understanding how analogue experiments differ from conventional experiments. Here (\S\ref{ssec:CE}), we argue that the crucial difference is the \textit{inaccessibility} of $T$, which prevents scientists from determining whether or not $T$ actually exemplifies the same type of behaviour as $S$ under the analogous conditions.

\subsection{Comparison with conventional experiments and simulations}\label{ssec:CE}

We take it that (conventional) experiments in science can be confirmatory; while the confirmatory status of analogue experiments is less clear (otherwise papers such as DTW and DHTW would not be written). Thus, in this subsection we attempt to distil the key differences between analogue and conventional experiments. We find that the key difference lies in the inaccessibility of the target system in an analogue experiment. This finding is consistent with DTW's characterisation as laid out above, as well as the use of the term by the other authors.

Two forms of experimental validity are standardly distinguished: an experiment is \textit{internally valid} if it reveals something about the particular system being used in the experiment, and an experiment is \textit{externally valid} if it is probative of the more general class of systems that the experimenters are interested in. If an experiment is conducted in order to learn only about the particular system being used, then the source and target systems are the same. We are not concerned with these kinds of experiments here, but instead with experiments where the source and target systems are different. 

In all cases of experiment in physics (apart from those being done in order to gain only internally valid results) the particular systems being used are meant to be representative of some more general class of systems which are supposed or conjectured, by current physics, to have in common some particular salient properties. Whether the experiment is being done in order to develop a hypothesis (model, modelling framework, or theory), or whether it is conducted in order to test some hypothesis, the salient properties of the system are those that feature in the developed, or tested, hypothesis. 

A first attempt to distinguish analogue from non-analogue experiments is then to say that, in \textit{non-analogue} experiments, $T$ is supposed to be---or is a potential candidate for being---of the same kind of system as $S$ \textit{according to current physics}; i.e., best current physics leads us to believe that the salient properties are strongly of the same kind in both systems. By contrast, one might imagine that, in \textit{analogue} experiments, $S$ and $T$ are not supposed to be the same kind of systems, but merely \textit{analogous} to one another, with $S$ exemplifying salient properties that are supposed to be only formally similar to those of $T$, according to their respective modelling frameworks.\footnote{This distinction is akin to one that has frequently been proposed to distinguish experiments from simulations, which holds that, in experiments, $S$ and $T$ bear material similarities, while in simulations, $S$ and $T$ bear only formal similarities. We agree with \cite{Winsberg2009, Winsberg2010} that a distinction along these lines is not tenable, whether between simulations and experiments, or analogue- and non-analogue experiments.} But---significantly---this first attempt at a definition will not do, because what it is to be `the same kind of system' in physics, for a specific application, is in fact determined by reference to the formal similarity of the modelling frameworks that describe the systems under the relevant conditions---any additional commonalities between $S$ and $T$ that are irrelevant to the phenomena of interest, are, by definition, not salient.

We mention a few examples to illustrate this. Firstly, consider an experiment on vortices in fluids: it may be performed on a single fluid, but is supposed to generalise to all fluids. What counts as `the same kind of system' in this case may just be any system that is described by the Navier-Stokes equation under the relevant conditions. The `same kind of system' may include air, water, smoke, liquid helium or something as exotic as a Bose-Einstein condensate---whatever system whose behaviour is supposed to be, according to current physics, described by the same modelling framework under the relevant conditions. Another example is that of phase transitions: here, the same kind of systems are those that share a critical exponent in the renormalisation group equations that describe their scaling behaviour. These systems may be incredibly diverse in constitution---metals, fluids, or even the universe itself (e.g., in cosmological domain wall formation)---yet, from the perspective of the phenomena of interest, they are of the same kind. In this case, the phenomenon is said to be \textit{universal}, and the systems that exemplify it are said to be in the same \textit{universality class}.\footnote{Cf. \citet{Batterman2000}.} Although these are technical terms typically used to refer to this sort of scaling behaviour in statistical mechanics, and related ideas in QFT, we henceforth co-opt them, and take them to refer more broadly to \textit{any systems that exemplify the same behaviour under some relevant conditions}; our doing so is in accordance with DTW's use of the term `universality'.

A final example is an experiment demonstrating resonance in a pendulum. In this case, a mass on a string can be treated as the same kind of system as an LC circuit---both exemplify the relevant behaviour and are, according to current physics, described by modelling frameworks that are syntactically isomorphic. On this view, two systems that are described by syntactically isomorphic modelling frameworks can be seen as the same kind of system at some higher level of abstraction, i.e., by appeal to more general features that the systems have in common, as captured by the the formal (structural) similarities of their descriptions. In the LC circuit and pendulum case, for instance, the class is that of \textit{simple harmonic oscillators}, while in the fluid case, it is that of (a specific kind of) \textit{fluid dynamics}, etc. Indeed, this move to a higher level of generality where physics can identify two compositionally distinct systems as being of the same kind is what is sought, by DTW, in arguing for the confirmatory nature of the dumb hole case. As discussed below (\S\ref{sec:why}), DTW's argument for the confirmatory nature of analogue experiments relies on the experimental demonstration of the robustness, or \textit{universality}, of the behaviour of interest in several systems of differing constitutions.

Now, all experiments (except those where internal validity is the only concern) rely on extrapolation in generalising their results to their target systems.\footnote{Admittedly, this has apparently been explored more in the context of biology, e.g., \citet[][\S3]{Baetu,Weber2018}, and in the literature on simulation, e.g., \citet{Winsberg2010}.} External validity is argued for in various ways, including via analogue reasoning. We thus suggest that the key difference between analogue and non-analogue experiments is not the exploitation of the syntactic isomorphism in Step 3 of DTW's characterisation.\footnote{Note that DTW do not themselves suggest that the syntactic isomorphism is the defining feature of analogue experiments, nor that its use is sufficient for distinguishing an analogue experiment from a conventional one.}

But then, given the fact that the dumb hole experiments---our exemplar of analogue experiments---aim to establish that, on some level of abstraction, black holes and dumb holes are in the same universality class\footnote{This is of course not the only aim or benefit of these experiments! But, as we argue in \S\ref{sec:why}, according to DTW, if dumb holes are to be confirmatory of black hole phenomena, it is precisely this universality that must be established.}, what can we say distinguishes an analogue experiment from other sorts of experiment in physics? We propose that the answer lies in the inaccessibility of the target system. Scientists cannot access $T$, and so they cannot confirm if $T$ and $S$ \textit{actually are} in the same universality class.

Although black holes are conjectured, by current physics, to be described by a particular modelling framework (QFT in curved spacetime, \S\ref{sec:HR})---it is not known if they are actually described by this framework, because we cannot access a black hole under the appropriate conditions. Thus, even though this framework is syntactically isomorphic to that of the dumb hole, we cannot establish that dumb holes and black holes really are the same kind of system. By contrast, physicists have access to pendulums and so can confirm that they are examples of simple harmonic oscillators under the relevant conditions, and thus can confirm that they are in the same universality class as the LC circuit under the appropriate conditions to demonstrate resonance. Hence, we suggest that the use of the term `analogue' to differentiate the dumb hole case from other examples of experiment in physics is an act of dignified restraint---it signifies recognition of our comparatively (severely) limited epistemic position with respect to the target system, compared with other cases of experiment.

Owing to this exceptionally limited epistemic position, together with the essential use of analogue reasoning that forms part of the definition of analogue experiment (as in, e.g., DTW's characterisation), analogue experiments retain the same status as other instances of analogue reasoning in science.\footnote{We of course acknowledge the difference between an argument and an experiment, but our point is that analogical arguments are still analogical arguments even when they relate to concrete, manipulable systems in the world.} Thus, the simple ability to perform experiments on one side of the analogy does not mean that the formal underpinnings of the analogy become any stronger. As concrete and manipulable as the accessible side of the analogy may be, it is the inferential step to the other side of the analogy which remains an instance of analogical reasoning. 

We close this section with the distinctions we propose, and which we use in the rest of the paper. In an...

\paragraph*{}
\textbf{Experiment}, $S$ and $T$ are---according to current physics---supposed to be, or are candidates for potentially being, \textit{the same kind of system for the purpose of interest} (i.e., as relevant to the phenomena under investigation); in other words, $T$ is supposed to be, or is conjectured to potentially be, in the same universality class as $S$ (on our broad interpretation of this term, as referring to systems described by syntactically isomorphic modelling frameworks) under the relevant conditions. (Note that $S$ and $T$ may be considered, by current physics, to be different systems at different levels of description, and/or under different conditions, e.g., they may have different microphysics that distinguishes them. The point is that these differences are irrelevant for the particular behaviour under investigation in the experiment, and so does not explicitly appear in the modelling frameworks describing the two systems).

\paragraph*{}
\textbf{Analogue experiment}, again, $S$ and $T$ are---according to current physics---supposed to be, or are candidates for potentially being, the same kind of system (i.e., in the same universality class, as defined by the relevant modelling framework) for the purpose of interest (this is because analogue experiments are a subclass of experiment). However, $T$ is inaccessible under the relevant conditions for confirming that it is actually the same kind of system as $S$ for the purpose of interest.\footnote{In a nutshell, this is also the reason why we will come to the conclusion in the following sections that dumb holes and other currently available analogue experiments on Hawking radiation simply cannot confirm the existence of gravitational Hawking radiation.}

\paragraph*{}
\textbf{Simulation}, $S$ is \textit{not} supposed---according to current physics---to be the same kind of system as $T$ for the purpose of interest (but to provide results that can potentially be translated into knowledge about $T$).

\paragraph*{}
  The idea is that what counts as the `same kind of system' is to some extent context and interest dependent. In this case, we are appealing to the interests of the researchers involved, and their employment of the best current physical theories (plus standard interpretations). When physicists refer to something as a `simulation' rather than an experiment, we take it that this signifies that $S$ is not supposed to be (for the purposes of interest) the same `kind' of system as $T$, and thus is not thought to be described by the modelling framework that defines the relevant universality class. Instead, perhaps, the system just produces output consistent with the modelling framework; it may, for instance, be thought to implement the modelling framework artificially, e.g., in a virtual environment (acknowledging of course that experiments and analogue experiments typically require a considerable amount of `setting up', and thus some degree of `artificiality'). Note that we include a definition of simulation only to make the contrast between simulation and experiment; we do not discuss simulation in this paper.\footnote{\citet{Winsberg2009, Winsberg2010} for instance characterises the distinction between simulation and experiment epistemically; for a simulation or experiment to be externally valid, $S$ (in either case) is hoped to ``stand in'' for $T$, by sharing formal descriptions, and in both cases this hope is justified by various background knowledge and assumptions of the researchers. In the case of experiment, this background is, for instance, the belief that $S$ and $T$ are the same kind of system, perhaps being materially similar. By contrast, in the case of simulation, the background is based on certain features of model building practices (background knowledge about model building practices is also used in the case of experiment, but according to Winsberg, the difference here is that in experiment this is mainly used to establish internal validity, while in the case of simulation it concerns external validity). Thus, on Winsberg's account of simulation, it is possible that $S$ and $T$ are supposed to be the same kind of system, but this supposition is not what is used to justify the belief in the external validity of the simulation.}

Thus, on our adopted distinctions, the dumb hole case is an example of an analogue experiment, rather than a simulation, contra the terminology used by DTW, which classifies it as `analogue simulation'. Another consequence of these definitions is that it is (in principle) possible that current analogue experiments are, in the future, recognised as conventional experiments. If scientists were to have sufficient access to black holes to confirm the relevant aspects of the modelling framework used to describe them---e.g., if direct detection methods for Hawking radiation were one day able to be developed, and these confirmed the existence of the phenomenon in black holes---then dumb holes could be used to experimentally probe black holes (given that the two would be recognised, on some higher level of abstraction, as `the same kind of system'), and such experiments could be confirmatory.

\section{Hawking radiation and dumb holes}\label{sec:HR}

This section lays bare the assumptions from which Hawking radiation is derived: the derivation relies on the framework of QFT in curved spacetime, and on a resolution of the so-called `trans-Planckian problem', which arises from the assumptions about trans-Planckian physics that that have to be fed into the theory. As Hawking radiation cannot be detected by conventional means, we briefly discuss analogous hydrodynamic Hawking effects.\footnote{Cf.\ also DTW (2017) and Th\'{e}bault (Forthcoming).}

\subsection{Hawking radiation}

In the mid-1970s, Hawking discovered that the QFT vacuum state becomes a state with (real) particles present if there is a black hole, i.e., that black holes \textit{radiate} \citep{hawking1975}. More precisely, he considered the vacuum state of a scalar field in the presence of an event horizon (that of the black hole) and found that it evolves into a thermal state of temperature $T$. Because the effect occurs due to the presence of a horizon, the temperature is associated with the black hole.\footnote{Hawking radiation has been derived for other kind of fields, including the electrodynamical vector field, although not necessarily using an analogous derivation. In fact, there seem to be at least five (putatively) independent derivations of the Hawking effect in the literature (cf.\ \cite{WallaceThermodynamics}).} It results in energy being radiated away, leading to the eventual `evaporation' of the black hole. The original calculation of the effect proceeds in the framework of QFT in curved spacetime.\footnote{It does not require a framework of semi-classical gravity in the sense of accounting for the backreaction of quantum matter to spacetime.}

A non-technical heuristic explanation attributes the effect to vacuum fluctuations leading to the separation of positive and negative modes through the black hole's event horizon: if such a pair forms near the event horizon, only one of the two partners may fall into the black hole, separating the pair and thus precluding their otherwise typical recombination, leaving the outward region with an effective excess energy. The escape of the excess particles to the asymptotic region is then what gives rise to outward radiation. Although this heuristic picture is quite imprecise---it does for instance not allow for any distinction between the Hawking effect and the Unruh effect (cf.\ \cite{barbado2016hawking})---it is sufficient for understanding the Hawking effect for what follows.

The Hawking radiation of typical astrophysical black holes has an extremely weak empirical signature, making its detection a near impossibility. As a black hole's temperature $T$ is proportional to its surface gravity $\kappa$---and, thereby, inversely proportional to its mass $M$---the temperature of the radiation becomes smaller the bigger the black hole considered. Consequently, black holes detectable in astronomical settings would emit radiation of an order several million times smaller than the temperature of the cosmic microwave background (\citeauthor{Thebault2016}, Forthcoming, 4). Additionally, black holes of a size sufficiently small that their putative radiation could perhaps be distinguished from noise have not yet been discovered (perhaps unsurprisingly, given that the smaller the black hole, the more rapidly it would supposedly evaporate away). But, as lab experiments on tailored artificial micro black holes seem impossible at the moment, the only options are to look out for small black holes either in particle detection facilities (see e.g., \cite{LargeHadronBH} or \cite{LargeHadronBH2} for proposals in this direction), or in the history of our universe. The Fermi-Gamma Ray Space Telescope (in a low-Earth orbit since 2008) has been searching, among other things, for radiative remnants of relatively small black hole objects from the beginning of the universe (see for instance, \cite{Fermi-Gamma}).

As stated, Hawking radiation is derived in the framework of QFT in curved spacetime; this is a non-trivial combination of GR and QFT which neglects the backreaction of matter on the curved spacetime. Such negligence is generally justifiable if the relevant length scales are large and curvature effects small. The detection of gravitational Hawking radiation would constitute a crucial confirmation of the applicability of the framework. So far, any available confirmation of QFT in curved spacetime is highly indirect in the sense that it does not even stem from very indirect measurements concerning predictions of particular QFTs in curved spacetime, but rather from claims that particular QFTs in curved spacetime obtain the right behaviour in \textit{various limits}\footnote{In other words, the exact same limit behaviour might be reproducible by myriad theories other than QFTs in curved spacetime.}: for instance, the geometric optics limit of electrodynamics in curved spacetime, which is a limit of \textit{quantum} electrodynamics in curved spacetime, describes light rays as tracing out null geodesics given sufficiently high frequency relative to the curvature scale. That light indeed moves on null geodesics in curved spacetime has been observationally verified, e.g. in gravitational lensing effects (see \cite{DysonGravitationalLensing}).

This evidence is not only indirect, but also highly unspecific about QFT in curved spacetime as it does not concern specific predictions of QFT in curved spacetime. In fact, it is currently only tenuously understood how well the theory applies empirically, i.e., whether its techniques of combining GR and QFT are in any sense valid. At the moment, there only seem to be two major applications and tests for QFT in curved spacetime that do concern specific predictions:

\begin{enumerate}
    \item The prediction of gravitational Hawking radiation.
    \item The prediction of a specific primordial density perturbation spectrum (associated with cosmic inflation scenarios\footnote{For an introduction to the topic of inflation, including the derivation of primordial density fluctuations, see e.g., \cite{Baumann}. For experimental results on the spectrum see \cite{PerturbationDensity}.}). 
\end{enumerate}

The status of Hawking radiation is, of course, open, whereas the specific primordial density spectrum has indeed been reported as successfully measured.\footnote{See for instance \cite{Lanusse}.} These tests can only be used to assess \textit{the conjunction} of QFT in curved spacetime \textit{and} certain other theories (mainly inflationary theories). This must strike one as problematic given that the empirical, conceptual or (even) the methodological status of (models of) inflation is disputed:

\begin{itemize}
    \item At the conceptual level, it is not clear that the issues that inflation addresses (horizon problem, flatness problems, fine-tuning) are problems in need of a solution, and even if they are, it may be that inflation just exchanges previous problems of fine-tuning by new ones (see \cite{McCoy}).
    \item At the empirical level, it seems as if the current predictions do include an accurate account of observed primordial density fluctuations---but again this does not mean that inflation is the only game in town. At best, the observation of the specific spectrum of fluctuations would only provide straightforward support to the QFT in curved spacetime framework \textit{provided that (some form of) inflation holds}. Two alternative models to inflation (matter bounce and string gas cosmology) arguably do equally well as inflationary models in predicting the mentioned fluctuations (see \cite{Brandenberger} for a discussion accessible to philosophers). Only provided that these models (also) use QFT in curved spacetime for deriving predictions of the density spectrum, the fluctuations observed could in principle also confirm QFT in virtue of one of these models (and not inflation) holding. And as a matter of fact, string gas cosmology, however, provides the same (currently measurable) predictions as inflation without any reference to QFT in curved spacetime. Instead this approach applies thermodynamic considerations of a gas made of strings (not particles).\footnote{Consider for instance \cite{Brandenberger} (p.\ 118) on this:

\begin{quote}
However, the physics of the generation mechanism is very different. In the case of inflationary cosmology, fluctuations are assumed to start as quantum vacuum perturbations because classical inhomogeneities are red-shifting. In contrast, in the Hagedorn phase of string gas cosmology there is no red-shifting of classical matter. Hence, it is the fluctuations in the classical matter which dominate. Since classical matter is a string gas, the dominant fluctuations are string thermodynamic fluctuations.
\end{quote}}
    \item At the methodological level, inflation models are accused of not even being falsifiable: there are so many models that surely some will give somewhat accurate predictions; worse, single inflation models in current inflationary theory typically involve multiverses so that within these multiverses surely one subsystem has the required properties to explain any desirable phenomena found in the cosmic microwave background---or so the common objection goes (see \cite{FactCheck}).
\end{itemize}

Thus, in light of inflation's controversial status, a more direct and less theory-dependent test of QFT in curved spacetime is needed. Unfortunately, such a test does not appear to be available among conventional experiments.

\subsection{Analogue Hawking radiation}

New hope of confirming the existence of Hawking radiation has emerged from analogue realisations of the gravitational Hawking effect through hydrodynamic structures (including Bose-Einstein condensates, BECs)\footnote{Bose-Einstein condensates can be treated at the level of a hydrodynamic approximation, see \cite{Garay}.}. In short, such hydrodynamic systems exhibit an effective background metric that is structurally identical to that of a spacetime with an event horizon. Then, once quantum fields enter on top, these systems exemplify an effect analogous to Hawking radiation.

To illustrate how the analogous Hawking effect is derived, consider a dumb hole, i.e., a hydrodynamic system with a local fluid stream running faster than the local speed of sound, with appropriate fluid density and velocity such that the acoustic horizon can be modelled by a Schwarzschild-like metric (in which acoustic waves play the same role as the matter fields do in Hawking's original derivation of the gravitational Hawking effect). Recall that\footnote{We are following \citeauthor{Thebault2016} (Forthcoming) here.} a classical fluid in the framework of continuum hydrodynamics is characterised by the mass density $\rho$, pressure $p$, and velocity $\vec{v}$ of the fluid volume element at each point of space. Only considering fluctuations around a fixed background configuration of the fluid, i.e. $(\vec{v}, p, \rho)=(\vec{v}_0, p_0, \rho_0)+(\vec{v}_1, p_1, \psi_1)$, one can linearise the fluid's continuity equation to arrive at 

\[\frac{1}{\sqrt{-g}} \frac{\partial}{\partial x_{\mu}} (\sqrt{-g} g^{\mu \nu} \frac{\partial}{\partial x^{\nu}}
 \psi_1 ) = 0,\]
through which the (density) fluctuations $\psi_1$ can be read as waves propagating in an effective background-spacetime structure $g^{\text{acoustic}}_{\mu \nu}$ determined by the fixed fluid background configuration $(\vec{v}_0, p_0, \rho_0)$. The background spacetime structure takes the form
 
 \[g^{\text{acoustic}}_{\mu \nu} = \frac{\rho_0}{c_{\text{sound}}}  \begin{pmatrix}
- (c^2_{\text{sound}} -v_0^2) & \vdots & -(v_0)_j \\
\dots & . & \dots \\
-(v_0)^i & \vdots & \delta_{ij}
\end{pmatrix},\]
where $c^2_{\text{sound}}$ denotes the speed of sound. Upon suitable choice of fluid density, velocity profile and fluid medium (sound speed), this acoustic metric can be put into a one-to-one relationship with a Schwarzschild-metric: 
 
 \[g^{\text{Schwarzschild}}_{\mu \nu} =   \begin{pmatrix}
- (c_0^2 - \frac{2 G M}{r}) & \vdots & -\sqrt{\frac{2GM}{r}} \vec{r_j} \\
\dots & . & \dots \\
-\sqrt{\frac{2GM}{r}} \vec{r_i} & \vdots & \delta_{ij}
\end{pmatrix} 
\]
for a given Schwarzschild mass $M$ and Newton constant $G$. Consequently, the wave equation for the fluctuation in the velocity potential in a suitably chosen fluid can be understood as analogous to a wave equation for a scalar field in a Schwarzschild spacetime. The field linked to these fluctuations can equally be quantised like the scalar field in a Schwarzschild spacetime. While the gravitational Hawking effect leads to radiation in the form of photons, the analogue hydrodynamical Hawking effects brings about the occurrence of phonons (`sound particles').

\cite{Steinhauer} claims to have experimentally verified the analogue (spontaneous) Hawking effect in a BEC. In Steinhauer's experiments, Hawking radiation is not directly measured; instead, Hawking radiation is linked to the entanglement of two particle partners coming out of vacuum fluctuations (consider the heuristic explanation above), and it is this entanglement which is measured (not the radiation itself). This becomes possible in the analogue scenario, as---unlike for a black hole---the measurement of correlations between entangled partners is straightforwardly possible as both sides of the horizon can be accessed by the observer in the lab.

\section{DTW's argument that analogue experiments can be confirmatory}\label{sec:why}

 According to \citeauthor{Thebault2016} (Forthcoming), an analogue experiment is capable of providing confirmation when we have reasons to believe in the external validity of the analogue experiment. Similarly, DTW argue that we can obtain `analogue confirmation' when we have (what they term) ``MEEGA'', or \textit{model-external, empirically-grounded arguments}, for the robustness of the modelling frameworks of both the analogue experiment and its target system. 

In the case of dumb holes as analogue experiments for Hawking radiation, Th\'{e}bault's arguments for the external validity of the dumb hole experiments, as well as DTW's MEEGA for the robustness of the two modelling frameworks are the same: it is the autonomy of the phenomena compared to microphysics---i.e., multiple realisability, or \textit{universality}. The universality of the Hawking effect in dumb holes has a theoretical as well as an experimental component. The theoretical universality is taken to be demonstrated by calculations such as those in \cite{UnruhSchutzhold}.\footnote{These calculations are further discussed in DTW and Th\'{e}bault regarding the trans-Planckian problem, but this is not important for our arguments.} In the actual black hole case, these are supposed to establish that---under certain assumptions on trans-Planckian physics (which we will list and discuss below)---the Hawking effect does not depend, to lowest order, on the details of the unknown theory of quantum gravity. In the dumb hole case, these arguments are supposed to establish that the existence of the analogue Hawking effect does not depend on the theory of the fluid at high energies where the assumption of continuity breaks down (i.e., the theory describing the fluid's constituents). The empirical aspect of DTW's MEEGA is supposed to come about once (or if) there are several experiments demonstrating the analogue effect in different sorts of systems. In such a case, then, according to DTW, we will have empirical confirmation of the existence of Hawking radiation in astrophysical black holes. 

The MEEGA, according to DTW (pp.\ 71, 73), are supposed to strengthen the analogy by virtue of the syntactic isomorphism between the modelling frameworks of the source and target systems, by providing a reason to believe in a relationship between the implicit assumptions used in the derivations of the phenomena in both of the frameworks. In the case where the MEEGA essentially turns on universality considerations, this means establishing that any differences between the source and target systems at high energy scales (which may or may not be captured by the analogy between these systems) are irrelevant for the appearance of the phenomena in question. Thus, the idea is that we can take an analogue experiment as capable of confirming claims about its target system when we have MEEGA establishing why any differences between the two systems do not affect the appearance of the behaviour of interest. And these universality arguments are not just theoretical, but have been experimentally confirmed through the appearance of the analogue phenomenon in various different source systems that are each supposedly analogous to the target system. In other words, using the language above (\S\ref{sec:analexp}), confirmation comes once we have amassed a collection of experimentally accessible systems within the same `universality class' that $T$ is supposed to be within.

Referring to DTW's argument structure quoted above (\S\ref{sec:analexp}, p.\ \pageref{def:ae}), the MEEGA are supposed to bolster Step 3, i.e., the analogy (syntactic isomorphism). It is this special MEEGA boost that is meant to elevate the analogy such that experimental observation of the analogue phenomena in the lab system can count as \textit{confirmation} of the existence (under the relevant conditions) of the actual phenomena in the target system.
\begin{quote}
On our view, confirmation of Hawking radiation via analogue simulation can only be established given the acceptance of a chain of reasoning involving universality arguments in combination with diverse realizations of the counterpart effect. These diverse realizations will thus simultaneously provide the empirical support for the MEEGA supporting the simulation, and realize the simulations themselves. (DTW, 2017, 76)
\end{quote}

Given the set-up above, two readings of an argument on analogue confirmation now suggest themselves:

On the first reading, the argument is basically as follows (subject to all the conditions cited in their Steps 1--5, \S\ref{sec:analexp}, p. \pageref{def:ae}, above): \textit{assuming that} the target system is described by $M_T$ (and $M_T$ is syntactically isomorphic to $M_S$, etc.), and the experiment on $S$ confirms that the microphysics of $S$ is irrelevant for the appearance of the phenomenon $P_S$, then, by analogy, we have confirmatory evidence \textit{that the microphysics of $T$ is also irrelevant to the appearance of $P_T$}. The experiment, in this case, is about demonstrating that the (precise form of) microphysics of various systems described by $M_S$ (i.e., the various systems experimented on) is irrelevant to the appearance of the phenomenon of interest in those systems. The problem with the argument on this interpretation is that it cannot be used to confirm that an unknown system actually displays the phenomenon of interest, but just that, \textit{if it did} display this phenomenon, its microphysics could be neglected in our description of this `macro' behaviour.

In order to actually have confirmation of the existence of $P_T$ by observation of $P_S$ (again, under all the conditions in Steps 1--5 of DTW's argument), we would need to confirm that $T$ is the same kind of system as $S$ for the purposes at hand---i.e., that they are described by syntactically isomorphic modelling frameworks under the relevant conditions. This is the second way of reading the argument of DTW, and the one most strongly suggested by their text\footnote{This reading is supported, e.g., by DTW's use of the idea of `universality', and statements like, ``we defend the claim that the phenomena of gravitational Hawking radiation could be confirmed in the case its counterpart is detected within experiments conducted on diverse realizations of the analogue model.'' \citet[55]{DardashtiQualitative}}: the empirically established universality arguments, i.e., the MEEGA, are supposed to provide evidence that $S$ and $T$ are in the same `universality class', and thus the appearance of $P_S$ could give us evidence that $P_T$ would also occur under the appropriate conditions. It is this interpretation of DTW's argument that we have adopted here (see, e.g., \S \ref{ssec:CE}).

On both readings, however, the problem we identify is the same---and it is independent of the MEEGA and Step 3---it is the assumption of Step 2: that $T$ is adequately described by $M_T$. Given that the target system is inaccessible in an analogue experiment, we cannot know if a particular modelling framework actually applies to it. Analogue experiments simply cannot tell us this, as we will elaborate on in the next section. 

\section{Analogue confirmation begs the question}\label{sec:critic}

\begin{figure}
\centering
\includegraphics[width=0.8\textwidth]{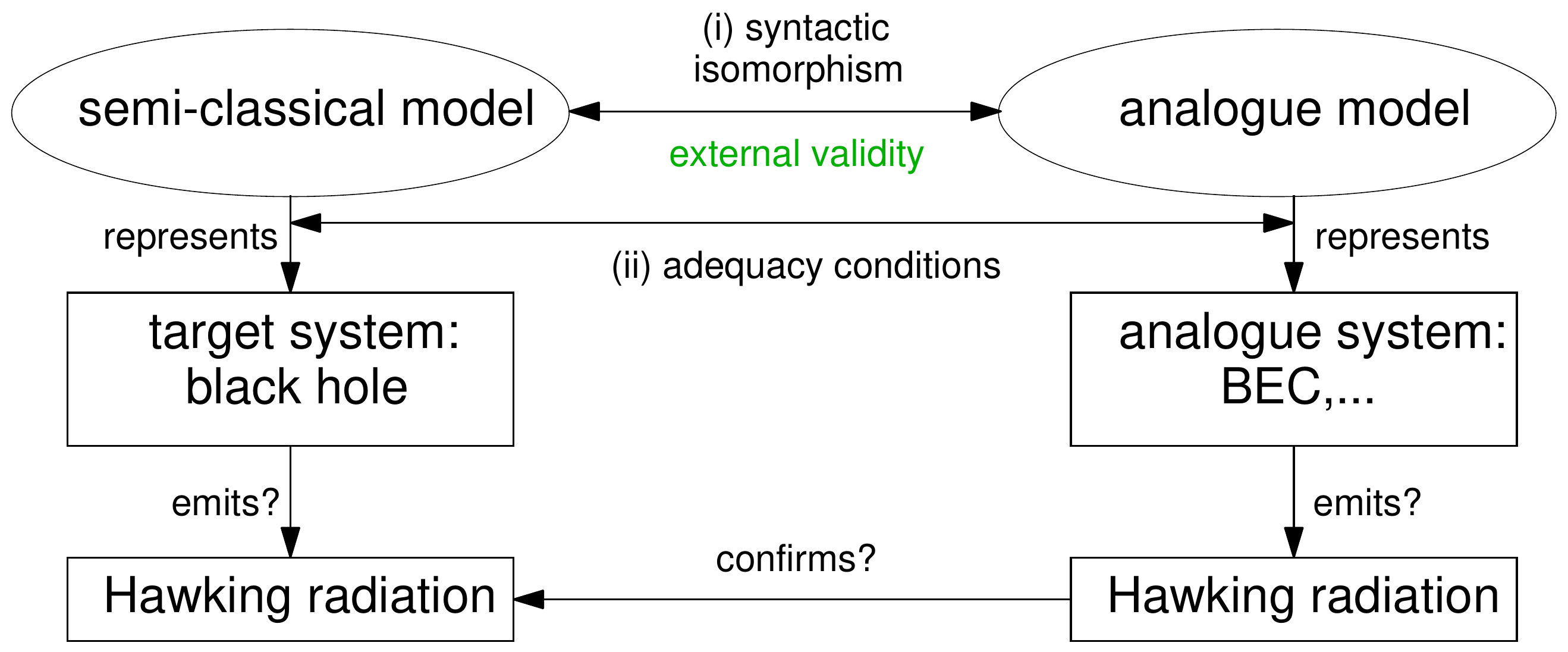}
\caption{The basic structure of the main argument of DTW (2017).}
\label{fig:analsimul}
\end{figure}

Let us review the structure of the main argument to the conclusion that analogue experiments can be confirmatory and that, in particular, analogue models such as dumb holes can confirm the existence of gravitational Hawking radiation in astrophysical black holes. This structure is depicted in Figure \ref{fig:analsimul}.
In this figure, the lab system is on the right side of the diagram, while the target system---the black hole---is on the left. We have our lab system, i.e., the analogue system, such as a Bose-Einstein condensate in the upper rectangular box. This analogue system is represented by an analogue model, i.e., by a theoretical description of the pertinent physics of the analogue system, and the experimental task is to detect possible (analogue) Hawking radiation that is emitted by the system. The corresponding target system, i.e., the black hole, is represented by the minimal semi-classical model used by Hawking, i.e., QFT on curved spacetime. The question we are ultimately interested in is, of course, whether there is (actual) Hawking radiation emitted from the target system, as was predicted by Hawking. 

According to this view, there now remain two main tasks to confirm gravitational Hawking radiation. First, there is the experimental work of setting up a suitable analogue system and detecting Hawking radiation emitted from such a system. Because Hawking radiation is weak, this task is far from trivial. As we have mentioned above, \cite{Steinhauer} claims to have successfully completed this first task for spontaneous analogue Hawking radiation. Although these results have not been reproduced so far, let us accept this claim for the sake of argument. After discharging this first task, there remains the second task: ascertaining the external validity of the analogue experiment---i.e., to extend its validity to include the target system, represented by the horizontal connections in Figure \ref{fig:analsimul}.

This second task divides into two subtasks, corresponding, essentially, to DTW's Step 3 on the one hand and to Steps 1 and 2 on the other. These two subtasks are represented in Figure \ref{fig:analsimul} by the top two horizontal connections between the two sides and are marked `(i)' and `(ii)', respectively. In order to establish the external validity of the analogue experiment, both of these subtasks must be completed. The first subtask, (i), establishes the formal analogy (the syntactic isomorphism) between the two modelling frameworks, i.e., between the descriptions of the analogue system and the target system. Just as in the case of the formal analogy between thermodynamic entropy and black hole entropy (or, more generally, between thermodynamic laws and black hole laws), the formal analogy between the modelling frameworks of Hawking radiation for the analogue and target systems must be supplemented by an argument establishing its physical salience.\footnote{Something Bekenstein failed to do, as argued in \citeauthor{Wuthrichbekenstein} (Forthcoming).} Specifically, the syntactic isomorphism between the models must be complemented with considerations establishing the \textit{physical adequacy} of the models in both cases.

If the physics of both systems is faithfully captured by formally analogous models, then there is reason to believe that both systems are governed essentially by the same physics, at least at the scale of interest (their microphysics will, in general, differ). The observation of Hawking radiation in one system would then confirm that there is Hawking radiation also in the other case. Hence, the second subtask, (ii), is of central importance to the success of DTW's argument. It is the task of Steps 1 and 2 to establish the physical adequacy of the modelling frameworks for the analogue and the target systems, respectively. Although this subtask consists in work proper to each side, evaluating different conditions for each---hence the separate Steps in DTW's argument---, and thus does not involve a direct link between the two sides of the analogy, the successful completion of this work on both sides results in a promotion of the formal analogy to one underwritten by physics (at the scale of interest). It is establishing this horizontal connection (ii) which guarantees the physical salience of the analogy and would, if borne out, justify the inferences from one physical system to the other.

The first part of (ii), DTW's Step 1, consists in work necessary to validate the physical adequacy of the analogue model. This side of the analogy is secured by a solid theoretical framework constituted by a tower of theories known to describe the physics of these analogue models at different levels of detail and universality. This tower includes, depending on the particular analogue system at stake, molecular hydrodynamics or Bogoliubov Bose-Einstein condensate theory---and, in any case, ultimately QFT. Not only are these analogue systems thus under firm theoretical control, these theories have independently been tested and confirmed on systems just like these analogue systems. Thus, the physical adequacy of the modelling framework on this side of the analogy is solidly underwritten by theory and experiment.

Given that Step 1 is thus unproblematic, the second part of (ii), Step 2 in DTW's argument, would ascertain, if successful, the physical adequacy of the semi-classical model of black holes. This is what makes Step 2 the keystone of the argument above. Specifically considering the case of Hawking radiation, DTW are assuming in Step 2 that black holes are accurately described by the modelling framework from which Hawking radiation is derived. In their own words:

\begin{quote}
    The system of type $T$ in this case is the astrophysical black hole. The modelling framework $M_T$ is a semi-classical model for gravity in which we have: (i) a fixed classical space-time that features the establishment of an event horizon via gravitational collapse leading to a black hole; and (ii) a quantum scalar field evaluated in the regions of past and future null infinity, which are assumed to be Minkowskian. The domain of conditions, $D_T$, is limited to the times after the collapse phase of the black hole, the details of which are assumed to be irrelevant. The work in dealing with the trans-Plankian problem discussed above is sufficient to demonstrate the viability of the modelling framework for some further refinement of $D_S$. For example, the conditions (i)–(iii) proposed by Unruh and Sch\"{u}tzhold [...] (DTW, 2017, 83).
\end{quote}
According to DTW, what would ultimately confirm gravitational Hawking radiation would be the existence of substantive reasons for believing that black holes fall into the same universality class as other systems that exhibit the phenomenon. Two generally necessary conditions for (re)producing Hawking radiation---and thus for establishing the external validity of the analogue experiment to the more general class of systems it is meant to represent---are, following \citet[\S5.1.1]{barceloliberativisser}, (a) that the form of the model is that of relativistic quantum fields on a classical effective background spacetime, and (b) that there is a horizon present in the geometry of the model. Furthermore, as DTW are aware (2017, 81) and discuss at length, for the universality claim made in \citet{UnruhSchutzhold} to extend to black holes, the so-called `trans-Planckian problem'\footnote{In the case of Hawking radiation (there are other trans-Planckian problems, e.g., the trans-Planckian problem in inflation), Unruh and Sch\"{u}tzhold characterise the issue as follows: ``in view of the (exponential) gravitational red-shift near the horizon, the outgoing particles of the Hawking radiation originate from modes with extremely large (e.g., trans-Planckian) wavenumbers. As the known equations of quantum fields in curved space-times are expected to break down at such wavenumbers, the derivation of the Hawking radiation has the flaw that it applies a theory beyond its region of validity. This observation poses the question of whether the Hawking effect is independent of Planckian physics or not." (p.\ 1)} must be solved in order to validate the semi-classical modelling framework assumed for the derivation of gravitational Hawking radiation. Unruh and Sch\"utzhold present a set of four\footnote{DTW (mistakenly) leave out the first of these four conditions (see quote of DTW above)---probably  because these are not made so explicit by Unruh and Sch\"{u}tzhold as are the other three conditions (referred to as (i)-(iii)).} jointly sufficient conditions on the high-energy behaviour to guarantee the universality of Hawking radiation: First, that the geometric optics only breaks down in the vicinity  of the horizon. Second, that there exists a privileged, freely falling frame. Third, that the Planckian modes start off in their ground state. Fourth, that the evolution of these modes is adiabatic. It should thus be clear that substantive assumptions must be in place for the semi-classical modelling framework assumed for the derivation of gravitational Hawking radiation to be physically adequate for black holes. 

In fact, the particular theoretical arguments for universality by \cite{UnruhSchutzhold} that DTW and Th\'{e}bault appeal to give us reason to doubt that they actually apply in the world. This is because these arguments rely on the assumption of Lorentz violation at the Planck scale\footnote{More precisely, Unruh and Sch\"utzhold's assumption of a (preferred) freely falling frame at high energies explicitly breaks (local) Lorentz invariance: \begin{quote}If we assume that the usual local Lorentz invariance is broken at the Planck scale via the introduction of preferred frames (where preferred frames are the frames in which Planckian physics displays maximal symmetry under time-inversion, for example) then the freely falling frame should be preferred (instead of the rest frame of the black hole, for example). (p.\ 9)\end{quote}}, and (conventional) experiments designed to test this claim have shown that it is unlikely to be true (except under strict conditions, see for instance, \cite{Mattingly}). Although this does not affect the arguments of DTW and Th\'{e}bault (since all they require is that there be \textit{some} theoretical and experimental arguments for universality, rather than these particular ones), it shows that any theoretical claims about black holes are fallible until faced with (conventional) experiment.

Our point is, then, simply, that it is not known if the particular modelling framework used in the derivation of Hawking radiation \textit{actually describes black holes in the first place}. This semi-classical modelling framework, $M_T$, with its particular assumptions and approximations, has neither been experimentally verified as actually applying to black holes nor is there a fundamental theory of quantum gravity which could theoretically underwrite its adequacy. That it has not been experimentally confirmed is what motivated analogue gravity to begin with. And the theoretical grasp on black hole physics is not remotely as firm as that of the analogue systems considered, e.g.\ terminating in a fundamental theory for which there is independent and substantive empirical confirmation. In fact, it is precisely in the hope to get some guidance in the search for such a theory that physicists turned to black hole physics. In sum, by assuming that black holes are accurately described by the modelling framework from which the derivation of Hawking radiation is a necessary consequence, DTW and Th\'{e}bault already assume the conclusion they are trying to establish---that Hawking radiation exists in black holes. It is in this sense that they are begging the question.\footnote{See Footnote \ref{fn:fallacy}.}

It seems to us that the literature on analogue experiment has lost focus on the kind of observational statement that is actually of interest here, which is the claim that,

\begin{itemize}
    \item[A.] Anything that is accurately described by our current black hole theory (i.e., QFT in curved spacetime)---if sufficiently realised in nature---will show multiply-realisable radiation, $X$. 
    \item[B.] Actual black holes (not some theoretical entities but genuine stuff in the world), show multiply-realisable radiation $X$.
\end{itemize}
Analogue experiments can only confirm A. It is then a further question whether actual black holes \textit{in the world} show the multiply-realisable effect $X$. In contrast, a conventional experiment could in principle establish B, which is the target hypothesis that we want confirmed in order to make progress on black hole physics and quantum gravity. 

The DTW paper remains at the qualitative level and does not develop its argument in the context of a particular account of confirmation. Our case so far only responds to this qualitative argument. According to DTW, any good confirmation framework \textit{should} be able to account for analogue experiments as a means of confirmation.\footnote{``From our perspective, if it proves that a philosophical model of confirmation cannot accommodate confirmation via analogue simulation at all, then this would be as much a problem for the model, as it would for analogue simulation.'' (DTW, 2017, 12).} They offer a Bayesian analysis of confirmation in the case of analogue experiments and prove theorems (in Bayesian confirmation theory) which they maintain establishes that we can have genuine confirmation through analogue experiments, i.e., that our rational credence in a hypothesis ought to strictly increase in the light of analogue evidence. In particular, on these theorems, analogue experiments in the lab \`a la Steinhauer can confirm the existence of Hawking radiation in astrophysical black holes. It is in this sense then that Bayesian analysis shows that analogue experiments not just provide some degree of confirmation, but can provide ``conclusive confirmatory support'': 
\begin{quote}
    As shown by a recent analysis in terms of Bayesian confirmation theory (Dardashti et al. 2016) [DHTW], given experimental demonstration of an array of analogue Hawking effects across a variety of different mediums the degree of confirmation conferred can be amplified very quickly. It is thus very plausible to think of analogue experiments prospective means for providing confirmatory support that is conclusive, rather than merely incremental. (Th\'{e}bault, \S1.2)
\end{quote}
Thus, the qualitative argument in DTW gets supplemented in DHTW by a `quantitative' one in terms of Bayesian confirmation theory. Let us analyse the quantitative argument.

\begin{figure}
\centering
\includegraphics[width=0.25\textwidth]{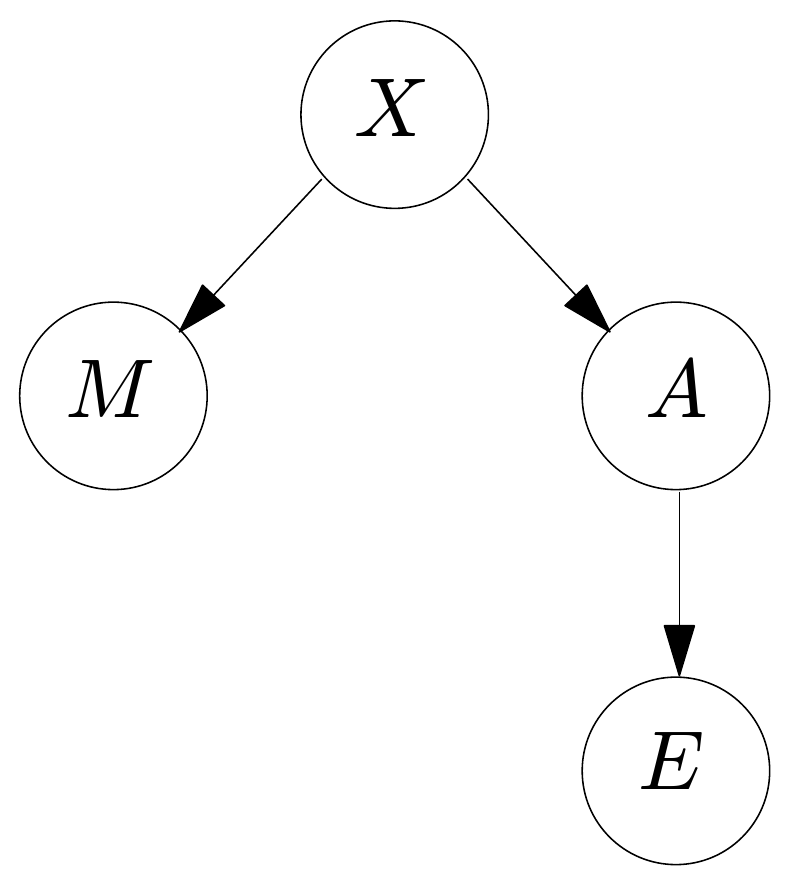}
\caption{The simplified Bayesian network of \citet[Figure 2]{DardashtiHawkingQuantitative}.}
\label{fig:network}
\end{figure}

The Bayesian analysis starts out by postulating a network representing the chain of inferences found in putative cases of confirmation through analogue experiments, shown in Figure \ref{fig:network}. A Bayesian network can be represented as a directed graph, with the variables as nodes and the probabilistic dependencies among them as directed edges. In Figure \ref{fig:network}, letters name binary variables capturing whether or not a corresponding proposition is true: $M$ related to the proposition that the modelling framework is an empirically adequate description of the physics of the target system at the scale of interest, $A$ that the modelling framework is an empirically adequate description of the physics of the analogue system at the scale of interest, $E$ that the empirical evidence obtains, i.e., that the phenomenon is observed, and $X$ that the universality arguments of the common background assumptions hold. It is these universality arguments, which are supposed to underwrite the external validity of the analogue experiments onto the target system. Thus, the $X$-factor connects the two sides of the analogy as a parent node to both $M$ and $A$, capturing the claim that, with some non-zero probability, both the target system and the analogue system fall under the same universality class. 

DHTW now prove a theorem (their Theorem 1) that, assuming a positive probabilistic dependence for each of the three edges in the network, $P(M|E) > P(M)$, i.e., that the probability that the modelling framework adequately represents the relevant physics of the target system increases if conditionalised on the evidence $E$. Among the three conditions expressing the positive probabilistic dependencies, one is of relevance to our concern: $P(M|X) > P(M|\neg X)$, which is equivalent to $P(X|M) > P(X)$. This is the condition which asserts the positive relevance of the universality arguments for the target system. Since this appears to be an epistemically appropriately modest demand given that we already assumed that $M$ probabilistic depends on $X$ when we drew the graph in Figure \ref{fig:network}, DHTW think that a rational agent should be compelled to accept it. 

Once we grant this assumption, it then follows that analogue evidence collected in the lab may increase the probability that the modelling framework adequately captures the relevant black hole physics, and, consequently, that black holes emit Hawking radiation. Thus, analogue experiments can putatively confirm our hypothesis concerning an inaccessible target system. But as we have tried to demonstrate, it is not at all clear why this presupposition should be granted: the circular dependence of conclusion on premise remains, as it must still be presupposed that black holes are the kind of system that, with at least some non-zero probability, exhibit certain physical behaviour, which is precisely what one seeks to establish with analogue confirmation. This is a weaker presupposition than just assuming, as the original argument by DTW did in Step 2, that the adequacy conditions are met in black holes: now we just posit that they possibly hold. In return, we also get much less than full confirmation: the only conclusion we can draw from Bayesian considerations is that the rational credence in the adequacy of the black hole modelling framework and hence in gravitational Hawking radiation ought to be increased given the observation of analogue Hawking radiation. Thus, this increase may be incremental, and far from providing ``conclusive confirmatory support''. 

In fact, it was thus already assumed that black holes at least probably fall under the relevant universality class when the network in Figure \ref{fig:network} was drawn: by connecting the variables $X$ and $M$, we thereby stipulated a correlation between black holes physics and the universality class. It is this step which presupposes (probabilistically) what we would like to establish, viz., that black holes fall into the relevant universality class. One may retort to this charge that all the theorem discussed above requires is that $P(M)\neq 0$ and that $P(M|X)$ is ever so slightly greater than $P(M|\neg X)$. Thus, it may be insisted, the initially required commitment to $M$ and to its probabilistic dependence on $X$ is minimal and so epistemically responsible. However, that these assumptions are not as innocent as they appear can be seen from the following straightforward consequence of the theorems in DHTW. Particularly by moving to multi-source confirmation \citep[\S4.2]{DardashtiHawkingQuantitative}, but in principle already in the simple case considered here (and in their \S4.1), we can perform a large number of terrestrial analogue experiments and thereby achieve an ever increasing (but in general bounded, cf.\ their \S4.2 and Appendix 2) degree of confirmation. Do enough lab work and we can be ever more certain that black holes emit Hawking radiation---without ever looking at a black hole! What we are interested in is precisely whether black holes are appropriately modelled by the model summarised in \S3.1. In the Bayesian framework, this translates into the question of whether an edge should be drawn between $X$ and $M$ in the Bayesian network in Figure \ref{fig:network} at all, and not just to what the priors are. 

Here is another way to think about what is going on, perhaps usefully illustrated with this example: although one can reasonably argue that a rational agent should assign non-zero probabilities (i.e. credence) to a universality thesis (and its negation), an example from the philosophy of mind shows what goes wrong in analogue confirmation. Suppose the functionalist assigns $P(U) = 0.75$ to the claim that any system with a certain functional design belongs to the universality class of consciousness and the dualist assigns $P(U) = 0.01$. So they will differ in priors, of course, but neither of them is permitted to choose either $0$ or $1$. Now we observe thousands of humans and each time, we raise the probability a bit---we confirm that the target system (such as the population of China as a whole) has consciousness. But of course the dualist wants to complain that no matter how many humans we are observing, we can never learn anything about the target system from that---in a sense observing humans does not address the research question.

At the end of the day, what is of course of interest is not just the confirmation of Hawking radiation as a multiply-realisable theoretical claim, but as a means of confirming more general frameworks leading to it. Whereas direct empirical confirmation of Hawking radiation also confirms the theoretical modelling framework, analogue confirmation cannot do so, as it already presupposes it. So, analogue confirmation cannot work due to its circularity. To spell this out a bit more: the only candidate for actual confirmation we have of the framework of QFT in curved spacetime so far seems to come from putative tests of (controversial) inflation, whereas a much more direct way of probing QFT in curved spacetime would indeed consist in testing gravitational Hawking radiation.\footnote{Even gravitational Hawking radiation cannot serve in a straightforward confirmation of QFT in a curved spacetime: as Hawking radiation can only be derived in a `QFT in curved spacetime' setting under further assumptions on the high-energy physics (to evade the prominent trans-Planckian problem), any test of it is in fact a test of both (1) the applicability of the QFT in curved spacetime framework, and (2) any proposed trans-Planckian physics behaviour used to evade the trans-Planckian problem (that the Hawking radiation needs to involve a cut-off of higher than Planck-scale energy behaviour), i.e., Hawking radiation can only serve as a test for both at the same time.}

To conclude: analogue confirmation does not work because the external validity of analogue experiments is unable to be determined. In spite of being unable to confirm claims about their target systems, however, these experiments are still scientifically useful. Foremost, analogue experiments that are \textit{internally valid} can be used to establish a concrete `proof of concept'. They facilitate the exploration of the modelling framework, and are heuristically profitable in that they can lead to discoveries about this framework that may otherwise have been obscure. Additionally, they can potentially demonstrate the robustness of the phenomena that they are supposed to be analogues of. But, we stress, that no matter how many analogue systems are built, the robustness demonstrated can not be confirmatory about claims regarding an unknown target system. 

Conversely, putting aside the issue of analogue confirmation, the hydrodynamic experiments (including, e.g., the dumb hole experiments) are of continued significance even if one is not interested in black holes at all (and even if these systems are not in fact in the same universality class as black holes). Shifting perspective, these experiments can instead be thought of as conventional experiments, with target systems that are not black holes, but rather other accessible (terrestrial) systems exhibiting the same behaviour---in this case, they are of interest in establishing the robustness of `universal Hawking radiation'. 


\section{Conclusion}\label{sec:conc}

A sufficient condition for an experiment or simulation to be potentially confirmatory is its being testable with respect to its target system. Material models and simulations, which (like analogue experiments) rely on analogy, can (we assume) be potentially confirmatory in the cases where they can be compared to their target system, and this testing can provide the arguments for their external validity. In those cases of material models and simulation where the target system is experimentally accessible, the fact that they can yield confirmation is reflective of the ability of the models and simulations to be tested (and not an indication that they have somehow transcended their categorisation as analogue models or simulations).

Analogue experiments, on the other hand, are, by definition, not testable with respect to their target systems: experiments can only be made on one side of the analogy. While DTW and Th\'{e}bault attempt to demonstrate that there may be other ways of establishing the external validity of an analogue experiment, these cannot overcome the key limitation on analogue experiments: the frontier of physics. For an analogue experiment to potentially confirm the existence of some phenomenon in an inaccessible system, on the basis of a formal analogy between the modelling frameworks $M_S$ and $M_T$ describing, respectively, the source and target systems (plus experimental observation of an analogous phenomenon in the source system under the relevant conditions, plus other MEEGA), it presupposes that $M_T$ actually describes the target system. But analogue experiments cannot tell us that $M_T$ actually describes the target system. And neither can arguments for multiple realisability, whether theoretical or empirical.

Owing to the tenuousness of the connection between dumb holes and black holes, the evidence that dumb hole experiments provide for the existence of Hawking radiation in black holes is not of a different epistemic sort than other instances of analogical reasoning. This means that if one believes that analogue experiments are capable of providing confirmation, then one should also be open to analogies as being able to increase credence in a hypothesis. Conversely, if one does not think that mere analogies yield confirmation, then neither do analogue experiments (after all, they necessarily involve analogical reasoning). We thus claim that analogue models of Hawking radiation are no more or less confirmatory than the analogy between thermodynamics and black hole thermodynamics, for example.

In short, assuming that analogue experiments cannot yield truly empirical confirmation, we argued that DTW beg the very question they set out to answer: the analogical reasoning necessary to get from an experimental result concerning an analogue system to a hypothesis regarding an inaccessible target system must presuppose that the formally same modelling framework applies to the inaccessible target system. This circularity is vicious given that analogue confirmation of gravitational Hawking radiation rests on the validity of QFT in curved spacetime, as well as the universality of the Hawking effect (i.e., its independence of high-energy effects), and that both of these hypotheses are to be tested by Hawking radiation in the first place. Thus, analogue confirmation does not seem to provide more than a consistency check for gravitational Hawking radiation. In other words, the reasoning involved in the theoretical argument around why gravitational black holes radiate can only show that \textit{if the formalism is indeed applicable to black holes} there would indeed be radiation. That the formalism is applicable to black holes---the real issue at stake---, can however not be settled by analogue confirmation, which leaves analogue confirmation, in the end, as an instance of analogue reasoning.

\bibliographystyle{newapa}

\bibliography{references}

\begin{thebibliography}{}

\bibitem[\protect\citeauthoryear{Baetu}{Baetu}{2016}]{Baetu}
Baetu, T.~M. (2016).
\newblock The `big picture': The problem of extrapolation in basic research.
\newblock {\em The British Journal for the Philosophy of Science}, {\em
  67\/}(4), 941--964.

\bibitem[\protect\citeauthoryear{Barbado, Barcel{\'o}, Garay \& Jannes}{Barbado
  et~al.}{2016}]{barbado2016hawking}
Barbado, L.~C., Barcel{\'o}, C., Garay, L.~J., \& Jannes, G. (2016).
\newblock Hawking versus {U}nruh effects, or the difficulty of slowly crossing
  a black hole horizon.
\newblock {\em Journal of High Energy Physics}, {\em 2016\/}(10), 161--174.

\bibitem[\protect\citeauthoryear{Barcel\'o, Liberati \& Visser}{Barcel\'o
  et~al.}{2011}]{barceloliberativisser}
Barcel\'o, C., Liberati, S., \& Visser, M. (2011).
\newblock Analogue gravity.
\newblock {\em Living Reviews of Relativity}, {\em 14}, 3.

\bibitem[\protect\citeauthoryear{Bartha}{Bartha}{2010}]{BarthaBook}
Bartha, P. (2010).
\newblock {\em By Parallel Reasoning: The Construction and Evaluation of
  Analogical Arguments}.
\newblock Oxford: Oxford University Press.

\bibitem[\protect\citeauthoryear{Bartha}{Bartha}{2016}]{sep-reasoning-analogy}
Bartha, P. (2016).
\newblock Analogy and analogical reasoning.
\newblock In E.~N. Zalta (Ed.), {\em The Stanford Encyclopedia of Philosophy\/}
  (Winter 2016 ed.). Metaphysics Research Lab, Stanford University.
\newblock
  \url{https://plato.stanford.edu/archives/win2016/entries/reasoning-analogy/}.

\bibitem[\protect\citeauthoryear{Batterman}{Batterman}{2000}]{Batterman2000}
Batterman, R.~W. (2000).
\newblock Multiple realizability and universality.
\newblock {\em The British Journal for the Philosophy of Science}, {\em
  51\/}(1), 115--145.

\bibitem[\protect\citeauthoryear{Baumann}{Baumann}{2011}]{Baumann}
Baumann, D. (2011).
\newblock {Inflation}.
\newblock In Csaki, C. \& Dodelson, S. (Eds.), {\em {Physics of the large and
  the small, TASI 09, proceedings of the Theoretical Advanced Study Institute
  in Elementary Particle Physics, Boulder, Colorado, USA, 1-26 June 2009}},
  (pp.\ 523--686).
\newblock \url{https://inspirehep.net/record/827549/files/arXiv:0907.5424.pdf}.

\bibitem[\protect\citeauthoryear{Bekenstein}{Bekenstein}{1972}]{bekenstein1972}
Bekenstein, J.~D. (1972).
\newblock Black holes and the second law.
\newblock {\em Lettere al Nuovo Cimento}, {\em 4}, 737--740.

\bibitem[\protect\citeauthoryear{Bekenstein}{Bekenstein}{1973}]{bekenstein1973}
Bekenstein, J.~D. (1973).
\newblock Black holes and entropy.
\newblock {\em Physical Review D}, {\em 7}, 2333--2346.

\bibitem[\protect\citeauthoryear{Brandenberger}{Brandenberger}{2014}]{Brandenberger}
Brandenberger, R. (2014).
\newblock Do we have a theory of early universe cosmology?
\newblock {\em Studies in History and Philosophy of Science Part B: Studies in
  History and Philosophy of Modern Physics}, {\em 46\/}(1), 109--121.

\bibitem[\protect\citeauthoryear{Crotty, Garcia-Bellido, Lesgourgues \&
  Riazuelo}{Crotty et~al.}{2003}]{PerturbationDensity}
Crotty, P., Garcia-Bellido, J., Lesgourgues, J., \& Riazuelo, A. (2003).
\newblock {Bounds on isocurvature perturbations from CMB and LSS data}.
\newblock {\em Physical Review Letters}, {\em 91}, 171301.

\bibitem[\protect\citeauthoryear{Dardashti, Hartmann, Th\'{e}bault \&
  Winsberg}{Dardashti et~al.}{2018}]{DardashtiHawkingQuantitative}
Dardashti, R., Hartmann, S., Th\'{e}bault, K.~P., \& Winsberg, E. (2018).
\newblock
\newblock Hawking radiation and analogue experiments: A {B}ayesian analysis.
\newblock \url{http://philsci-archive.pitt.edu/14819/}.

\bibitem[\protect\citeauthoryear{Dardashti, Th\'{e}bault \& Winsberg}{Dardashti
  et~al.}{2017}]{DardashtiQualitative}
Dardashti, R., Th\'{e}bault, K.~P., \& Winsberg, E. (2017).
\newblock Confirmation via analogue simulation: What dumb holes could tell us
  about gravity.
\newblock {\em The British Journal for the Philosophy of Science}, {\em 68},
  55--89.

\bibitem[\protect\citeauthoryear{Dimopoulos \& Landsberg}{Dimopoulos \&
  Landsberg}{2001}]{LargeHadronBH2}
Dimopoulos, S. \& Landsberg, G. (2001).
\newblock Black holes at the large hadron collider.
\newblock {\em Physical Review Letters}, {\em 87}, 161602.
\newblock \url{https://link.aps.org/doi/10.1103/PhysRevLett.87.161602}.

\bibitem[\protect\citeauthoryear{Dougherty \& Callender}{Dougherty \&
  Callender}{ming}]{DoughertyCallender}
Dougherty, J. \& Callender, C. (Forthcoming).
\newblock Black hole thermodynamics: More than an analogy?
\newblock In B.~Loewer (Ed.), {\em Philosophy of Cosmology}. Cambridge
  University Press.
\newblock \url{http://philsci-archive.pitt.edu/13195/}.

\bibitem[\protect\citeauthoryear{{Dyson}, {Eddington} \& {Davidson}}{{Dyson}
  et~al.}{1920}]{DysonGravitationalLensing}
{Dyson}, F.~W., {Eddington}, A.~S., \& {Davidson}, C. (1920).
\newblock {A Determination of the Deflection of Light by the Sun's
  Gravitational Field, from Observations Made at the Total Eclipse of May 29,
  1919}.
\newblock {\em Philosophical Transactions of the Royal Society of London Series
  A}, {\em 220}, 291--333.

\bibitem[\protect\citeauthoryear{Euv{\'e}, Michel, Parentani, Philbin \&
  Rousseaux}{Euv{\'e} et~al.}{2016}]{French}
Euv{\'e}, L.-P., Michel, F., Parentani, R., Philbin, T.~G., \& Rousseaux, G.
  (2016).
\newblock Observation of noise correlated by the hawking effect in a water
  tank.
\newblock {\em Physical review letters}, {\em 117\/}(12), 121301.

\bibitem[\protect\citeauthoryear{Garay, Anglin, Cirac \& Zoller}{Garay
  et~al.}{2000}]{Garay}
Garay, L.~J., Anglin, J., Cirac, J.~I., \& Zoller, P. (2000).
\newblock Sonic analog of gravitational black holes in bose-einstein
  condensates.
\newblock {\em Physical Review Letters}, {\em 85\/}(22), 4643.

\bibitem[\protect\citeauthoryear{Giddings \& Thomas}{Giddings \&
  Thomas}{2002}]{LargeHadronBH}
Giddings, S.~B. \& Thomas, S. (2002).
\newblock High energy colliders as black hole factories: The end of short
  distance physics.
\newblock {\em Physical Review D}, {\em 65}, 056010.

\bibitem[\protect\citeauthoryear{Hawking}{Hawking}{1975}]{hawking1975}
Hawking, S.~W. (1975).
\newblock Particle creation by black holes.
\newblock {\em Communications in Mathematical Physics}, {\em 43}, 199--220.

\bibitem[\protect\citeauthoryear{Hempel}{Hempel}{1965}]{Hempel1965}
Hempel, C.~G. (1965).
\newblock {\em Aspects of Scientific Explanation and Other Essays in the
  Philosophy of Science}.
\newblock New York: Free Press.

\bibitem[\protect\citeauthoryear{Hesse}{Hesse}{1964}]{Hesse1964}
Hesse, M. (1964).
\newblock Analogy and confirmation theory.
\newblock {\em Philosophy of Science}, {\em 31\/}(4), 319--327.

\bibitem[\protect\citeauthoryear{Ijjas, Steinhardt \& Loeb}{Ijjas
  et~al.}{2017}]{FactCheck}
Ijjas, A., Steinhardt, P., \& Loeb, A. (2017).
\newblock Pop goes the universe.
\newblock last checked: 11 March 2018.

\bibitem[\protect\citeauthoryear{Kiefer}{Kiefer}{2007}]{KieferBook}
Kiefer, C. (2007).
\newblock Why quantum gravity?
\newblock In I.-O. Stamatescu \& E.~Seiler (Eds.), {\em Approaches to
  Fundamental Physics}  (pp.\ 123--130). Springer.

\bibitem[\protect\citeauthoryear{Lanusse, Paykari, Starck, Sureau \&
  Bobin}{Lanusse et~al.}{2014}]{Lanusse}
Lanusse, F., Paykari, P., Starck, J.~L., Sureau, F., \& Bobin, J. (2014).
\newblock {PRISM: Recovery of the primordial spectrum from Planck data}.
\newblock {\em Astronomy and Astrophysics}, {\em 571}, L1.

\bibitem[\protect\citeauthoryear{Mattingly}{Mattingly}{2005}]{Mattingly}
Mattingly, D. (2005).
\newblock Modern tests of {L}orentz invariance.
\newblock {\em Living Reviews in Relativity}, {\em 8\/}(1), 5.

\bibitem[\protect\citeauthoryear{McCoy}{McCoy}{2015}]{McCoy}
McCoy, C. (2015).
\newblock Does inflation solve the hot big bang model's fine-tuning problems?
\newblock {\em Studies in History and Philosophy of Science Part B: Studies in
  History and Philosophy of Modern Physics}, {\em 51}, 23--36.

\bibitem[\protect\citeauthoryear{Rousseaux, Ma{\"\i}ssa, Mathis, Coullet,
  Philbin \& Leonhardt}{Rousseaux et~al.}{2010}]{Rousseaux2010horizon}
Rousseaux, G., Ma{\"\i}ssa, P., Mathis, C., Coullet, P., Philbin, T.~G., \&
  Leonhardt, U. (2010).
\newblock Horizon effects with surface waves on moving water.
\newblock {\em New Journal of Physics}, {\em 12\/}(9), 095018.

\bibitem[\protect\citeauthoryear{Rousseaux, Mathis, Ma{\"\i}ssa, Philbin \&
  Leonhardt}{Rousseaux et~al.}{2008}]{rousseaux2008observation}
Rousseaux, G., Mathis, C., Ma{\"\i}ssa, P., Philbin, T.~G., \& Leonhardt, U.
  (2008).
\newblock Observation of negative-frequency waves in a water tank: a classical
  analogue to the hawking effect?
\newblock {\em New Journal of Physics}, {\em 10\/}(5), 053015.

\bibitem[\protect\citeauthoryear{Steinhauer}{Steinhauer}{2016}]{Steinhauer}
Steinhauer, J. (2016).
\newblock Observation of quantum {H}awking radiation and its entanglement in an
  analogue black hole.
\newblock {\em Nature Physics}, {\em 12\/}(10), 959--965.

\bibitem[\protect\citeauthoryear{Th\'{e}bault}{Th\'{e}bault}{ming}]{Thebault2016}
Th\'{e}bault, K.~P. (Forthcoming).
\newblock What can we learn from analogue experiments?
\newblock In R.~Dawid, R.~Dardashti, \& K.~Th\'{e}bault (Eds.), {\em
  Epistemology of Fundamental Physics}. Cambridge: Cambridge University Press.
\newblock arXiv:1610.05028.

\bibitem[\protect\citeauthoryear{Torres, Patrick, Coutant, Richartz, Tedford \&
  Weinfurtner}{Torres et~al.}{2017}]{superradiance}
Torres, T., Patrick, S., Coutant, A., Richartz, M., Tedford, E.~W., \&
  Weinfurtner, S. (2017).
\newblock Rotational superradiant scattering in a vortex flow.
\newblock {\em Nature Physics}, (13), 833--836.

\bibitem[\protect\citeauthoryear{Ukwatta, MacGibbon, Parke, Dhuga, Rhodes,
  Eskandarian, Gehrels, Maximon \& Morris}{Ukwatta et~al.}{2010}]{Fermi-Gamma}
Ukwatta, T.~N., MacGibbon, J.~H., Parke, W.~C., Dhuga, K.~S., Rhodes, S.,
  Eskandarian, A., Gehrels, N., Maximon, L., \& Morris, D.~C. (2010).
\newblock {Sensitivity of the FERMI Detectors to Gamma-Ray Bursts from
  Evaporating Primordial Black Holes (PBHs)}.
\newblock In {\em {On recent developments in theoretical and experimental
  general relativity, astrophysics and relativistic field theories.
  Proceedings, 12th Marcel Grossmann Meeting on General Relativity, Paris,
  France, July 12-18, 2009. Vol. 1-3}}, (pp.\ 1588--1590).

\bibitem[\protect\citeauthoryear{Unruh \& Sch\"{u}tzhold}{Unruh \&
  Sch\"{u}tzhold}{2005}]{UnruhSchutzhold}
Unruh, W. \& Sch\"{u}tzhold, R. (2005).
\newblock Universality of the {H}awking effect.
\newblock {\em Physical Review D}, {\em 71\/}(2), 024028.

\bibitem[\protect\citeauthoryear{Unruh}{Unruh}{1981}]{unruh1981}
Unruh, W.~G. (1981).
\newblock Experiments with black hole evaporation?
\newblock {\em Physical Review Letters}, {\em 46}, 1351--1353.

\bibitem[\protect\citeauthoryear{Wallace}{Wallace}{2017}]{WallaceThermodynamics}
Wallace, D. (2017).
\newblock The case for black hole thermodynamics, part i: phenomenological
  thermodynamics.
\newblock {\em arXiv preprint arXiv:1710.02724}.

\bibitem[\protect\citeauthoryear{Weber}{Weber}{2018}]{Weber2018}
Weber, M. (2018).
\newblock Experiment in biology.
\newblock In {\em The Stanford Encyclopedia of Philosophy\/} (Summer ed.).
  https://plato.stanford.edu/archives/sum2018/entries/biology-experiment/.

\bibitem[\protect\citeauthoryear{Weinfurtner, Tedford, Penrice, Unruh \&
  Lawrence}{Weinfurtner et~al.}{2013}]{Weinfurtner}
Weinfurtner, S., Tedford, E.~W., Penrice, M.~C., Unruh, W.~G., \& Lawrence,
  G.~A. (2013).
\newblock Classical aspects of hawking radiation verified in analogue gravity
  experiment.
\newblock In {\em Analogue Gravity Phenomenology}  (pp.\ 167--180). Springer.

\bibitem[\protect\citeauthoryear{Winsberg}{Winsberg}{2009}]{Winsberg2009}
Winsberg, E. (2009).
\newblock A tale of two methods.
\newblock {\em Synthese}, {\em 169\/}(3), 575--592.

\bibitem[\protect\citeauthoryear{Winsberg}{Winsberg}{2010}]{Winsberg2010}
Winsberg, E. (2010).
\newblock {\em Science in the Age of Computer Simulation}.
\newblock University of Chicago Press.

\bibitem[\protect\citeauthoryear{W{\"u}thrich}{W{\"u}thrich}{2018}]{Wuthrichbekenstein}
W{\"u}thrich, C. (2018).
\newblock Are black holes about information?
\newblock In R.~Dawid, R.~Dardashti, \& K.~Th\'ebault (Eds.), {\em Why Trust a
  Theory? Epistemology of Fundamental Physics}  (pp.\ 202--223). Cambridge:
  Cambridge University Press.

\end{thebibliography}

\end{document}